\documentclass[a4paper,superscriptaddress,aps,prb,twocolumn,floatfix,citeautoscript]{revtex4}

\makeatletter
\def\@dotsep{4.5}
\makeatother

\usepackage{amsmath,amsbsy,amsfonts,mathrsfs,graphicx,sidecap,bm}
\usepackage{dcolumn}

\newcommand{\I}{\mathrm{i}}
\newcommand{\D}{\mathrm{d}}

\renewcommand{\Im}{\mathfrak{Im}\,}

\newcommand{\mint}[1]{\int\! \D^{3} #1 \, }

\newcommand{\ba}{\begin{array}}
\newcommand{\ea}{\end{array}}

\newcommand {\lsi}{Laboratoire des Solides Irradi\'es,
  CNRS-CEA-\'Ecole Polytechnique, Palaiseau, France}

\newcommand {\coimbra}{Centro de F\'{\i}sica Computacional, Departamento de F{\'\i}sica, Universidade de
  Coimbra, Coimbra, Portugal}

\newcommand{\berlin}{Institut f{\"{u}}r Theoretische Physik, Fachbereich Physik der
Freie Universit{\"{a}}t Berlin, Arnimallee 14, D-14195 Berlin, Germany}

\newcommand {\sanse}{European Theoretical Spectroscopy Facility, Departamento  de F{\'{\i}}sica de Materiales, Universidad del Pa{\'{i}}s Vasco,  Centro Mixto CSIC-UPV, and Donostia International Physics Center  (DIPC), Av. Tolosa 72, E-20018 San Sebasti\'an,  Spain}  

\newcommand {\lyon}{LPMCN, Universit\'e Claude Bernard Lyon I and CNRS UMR 5586, 69622 Villeurbanne, France}

\newcommand {\etsf}{European Theoretical Spectroscopy Facility}

\begin{document}

\title{Cluster-surface and cluster-cluster interactions: ab initio calculations and modelling of asymptotic van der Waals forces}

\date{\today}

\author{Silvana Botti}
\email{silvana.botti@polytechnique.edu}
\affiliation{\lsi}
\affiliation{\coimbra}
\affiliation{\lyon}
\affiliation{\etsf}

\author{Alberto Castro}
\affiliation{\berlin}
\affiliation{\etsf}

\author{Xavier Andrade}
\affiliation{\sanse}
\affiliation{\etsf}

\author{Angel Rubio}
\affiliation{\sanse}
\affiliation{\etsf}

\author{Miguel A. L. Marques}
\affiliation{\lyon}
\affiliation{\coimbra}
\affiliation{\etsf}

\begin{abstract}
  We present fully ab-initio calculations of van der Waals
  coefficients for two different situations: i)~the interaction
  between hydrogenated silicon clusters; and ii)~the interactions
  between these nanostructures and a non metallic surface (a silicon
  or a silicon carbide surface). The methods used are very efficient,
  and allow the calculation of systems containing hundreds of atoms.
  The results obtained are further analyzed and understood with the
  help of simple models. These models can be of interest for molecular
  dynamics simulations of silicon nanostructures on surfaces, where
  they can give a very fast yet sufficiently accurate determination of
  the van der Waals interaction at large separations.
\end{abstract}

\maketitle

\section{Introduction}

The van der Waals interaction is a common presence in the worlds of
Physics, Chemistry, and Biology~\cite{vanderwaals}. Studied for more
than a century~\cite{vanderwaals-historical}, it is a dispersive force
that, being weak at short distances, becomes the dominant attraction
between neutral bodies at large separations. It results from the
non-zero multipole-multipole attraction stemming from transient
quantum fluctuations. It is the interplay between the electrostatic
and dispersive interactions that determines many interesting phenomena
in Nature, even in the macroscopic world. For example, it is van der
Waals interactions that are responsible for the remarkable ability of
geckos to hold to surfaces~\cite{tian-2006}.  A significant example of
application of van der Waals interactions comes from some operating modes of
atomic force microscopes~\cite{binnig-1986}.

However, the realm of van der Waals forces, being quantum in nature,
is the world of atoms and molecules --- the nano world. In fact, these
forces determine the structure of DNA molecules, the folding and
dynamics of proteins, the adsorption of atoms, molecules or
nanostructures on surfaces, etc.  Moreover, the van der Waals
atom-surface interaction has also been recently studied due to their
influence on the quantum reflection of ultracold atoms on
surfaces~\cite{shimizu-2001}. In fact, the upsurge of interest on
Bose-Einstein condensation of ultracold atoms confined in magnetic
traps~\cite{harber-2003} near a surface should fuel research on
atom-surface dispersion interactions, since they are key factors for
the stability of the condensate. They are also key ingredients in the
building and functioning of many of the systems relevant for the
emerging fields of nanotechnology and biotechnology. For example,
their effect was shown to have a profound influence on the oscillatory
behavior of microstructures when surfaces are in close proximity (100
nm)~\cite{chan-2001}.

In this context, the purpose of this Article is to present fully
ab-initio calculations of the van der Waals coefficients for the
interaction between nanostructures (silicon clusters, in particular)
and between these nanostructures and surfaces. These results are then
used to formulate simple models that describe with enough precision
the van der Waals interactions. The models are important, as they can
be the starting point for, e.g., molecular dynamics simulations of the
behavior of silicon nanostructures at surfaces.
We will be looking at large separations, when the overlap between
the electronic clouds is negligible. At shorter distances, the situation
is considerably more complicated and no satisfying description has emerged yet.

Typically, van der Waals interactions decay with an inverse power of
the distance between the two bodies under consideration; The exponent
depends on their their dimensionality or their metallic character
\cite{dobson-2006}. We are interested in two specific cases:

A)~{\em The interaction between two finite nanostructures}, namely
atomic clusters of silicon, saturated with hydrogen. We restrict
ourselves to the more interesting non-retarded regime, i.e. when the
time that it takes for the photons to travel between the two molecules
is negligible. In this case, the van der Waals interaction energy has
an expansion with respect to the inverse of the intermolecular
distance ($1/R$) of the form
\begin{equation}
  \label{eq:eofr} \Delta E(R) = -\sum_{n=6}^{\infty}
  \frac{C_n}{R^{n}}\,
\end{equation}
in terms of the Hamaker constants $C_n$~\cite{hamaker-1937, vanderwaals}. The first term $C_6$ for a pair of
molecules $A$ and $B$, averaged over all possible orientations, can be
obtained through the relation (atomic units
will be used hereafter):
\begin{equation}
  \label{eq:c6} C^{AB}_6 = \frac{3}{\pi}\int_0^{\infty} \!\!\!{\textrm
  d}u \;\alpha^{(A)}({\textrm i}u)\;\alpha^{(B)}({\textrm i}u)\,,
\end{equation}
where $\alpha^{(X)}({\textrm i}u)$ is the average of the dipole
polarizability tensor of molecule $X$, $\boldsymbol{\alpha}^{(X)}$,
evaluated at the imaginary frequency ${\textrm i}u$:
\begin{equation}
  \alpha^{(X)}({\textrm i}u) = \frac{1}{3} {\textrm Tr}
  [\boldsymbol{\alpha}^{(X)}({\textrm i}u)]\,.
\end{equation}
The higher order terms in the expansion (\ref{eq:eofr}) can in a
similar way be written in terms of higher order polarizability
tensors. For example, the $C_8$ coefficient will depend on the
dipole-quadrupole dynamic polarizability (see, e.g.,
Ref.~~\onlinecite{jeziorsky94}). In this article we will focus on the
leading term of the expansion, i.e. the Hamaker constant $C_6$.

B)~{\em The interaction between a nanostructure and a surface}. Here
we will focus on silicon and silicon carbide surfaces.  In this case,
the leading term of the expansion of the van der Waals energy as a
function of the distance $Z$ between the cluster and the surface is
proportional to $Z^{-3}$. This fact was first established by
Lennard-Jones~\cite{lennard-1932}, who used a model with a perfectly
reflecting metal. The theory was later developed by Casimir and
Polder~\cite{casimir-1946}, and by Lifschitz~\cite{Lifshitz56}. For a
wide range of particle-substrate distances (from approximately 1~nm to
even 10$^3$~nm), the interaction energy is given by:
\begin{equation}
  \Delta E(Z) = -\frac{C_3}{(Z-Z_0)^3}\,,
\end{equation}
where $Z_0$ is a ``reference plane'', and $C_3$ is the Lifshitz
coefficient. This coefficient can be calculated from the dynamical
polarizability of the cluster, $\alpha(\I u)$, and the macroscopic
dielectric function of the bulk material, $\epsilon_{\textrm M}({\textrm i}u)$,
both evaluated at imaginary frequencies, ${\textrm i}u$:
\begin{equation}
  \label{eq:c3} C_3 = \frac{1}{4 \pi} \int_0^{\infty} \!\!\!{\textrm
  d}u \;\alpha({\textrm i}u) \frac{\epsilon_{\textrm M}({\textrm i}u)
  -1}{\epsilon_{\textrm M}({\textrm i}u) +1} \,.
\end{equation}

Note that $C_3$ is expressed only in terms of quantities calculated
for the bulk crystal. This expression is a general result, also valid
for metallic surfaces. The quantity that depends on the
characteristics of the surface is the position of the reference plane
$Z_0$. However, for semiconducting surfaces, it can be
shown~\cite{zaremba76} that, in absence of local field effects, $Z_0$
is equal to $a/2$, where $a$ is the interplanar distance. Moreover, it
is known that even relatively large local field corrections give rise
to rather small shifts of the reference plane~\cite{zaremba76}.  Note,
however, the position of the reference plane $Z_0$ is a more delicate
issue in the case of a metal, as positioning the reference plane at a
distance of $a/2$ from the surface can lead to significant errors in
the interaction energy (i.e. about 30\% for a noble metal
surface~\cite{zaremba76}). Further analysis to determine the dependence of
van der Waals interactions on the surface response are under progress.

It is interesting to remark that it took a long time to verify
experimentally the predicted $Z^{-3}$ dependence. Although this term
dominates a very wide distance region, in the short distance regime it
is only a tail of the particle potential whose minimum determines the
adsorption. The first precise measurement of the van der Waals
coupling between an atom and a surface was reported in
Ref.~~\onlinecite{sandoghdar-1992}; later on some more experiments
have followed~\cite{fichet-2007}. The experimental difficulties are,
in fact, in pair with the theoretical ones, since fully ab initio
calculations are also challenging. Most of the calculations reported
in the literature~\cite{caride05, patil-2002} are generally limited to
atoms or very small molecules, and the bulk detailed microscopic
structure is replaced by some model (e.g., the stabilized jellium
model for metals). An interesting approach consists in modelling
the molecular polarizability using the form of the long-wavelength density
response of a homogeneous electron gas. The van der Waals interaction is thus efficiently
described by the ground-state electron densities of the interacting species,
obtaining $C_6$ coefficients that on average deviate 9\% from those
obtained by TDDFT calculations\cite{andersson-1996}. All approximations
were justified by the authors by the difficulty of treating medium size molecules within a 
full ab initio TDDFT approach.

It is the purpose of our work to demonstrate that
current ab initio techniques permit the calculation of  the van der Waals
coefficients for large nanostructures interacting with realistically
described surfaces. The rest of this Article is structured as follows: In
Section~\ref{sec:methods} we review the methods used to evaluate the
dynamical polarizabilities and the dielectric functions at imaginary
frequency; in the following section we present the results of our ab
initio calculations, that are then further analyzed in
Section~\ref{sec:models} using some simple models. Finally, we draw
some conclusions in Section~\ref{sec:conclusions}.

\section{Methods}
\label{sec:methods}

The main ingredients to evaluate the van der Waals coefficients are
therefore the electronic polarizability $\alpha$ of the cluster and
the dielectric constant of the bulk material $\varepsilon$, both
evaluated at imaginary frequencies. The computational methods and the
problems involved in the calculation of these two quantities are quite
different, so we will discuss them separately.

\subsection{Dynamical polarizabilities}

In principle, one can obtain the dynamical polarizabilities by making
use of any quantum-chemistry theory capable of handling time-dependent
perturbations. As the nanostructures we are interested in can be
fairly large, we choose the time-dependent (TD) extension of density
functional theory (DFT), since this approach provides an excellent
compromise between accuracy and feasibility.  During the past decade,
TDDFT~\cite{tddft-book} has become one of the most important tools to
study electronic excitations of molecular systems, especially for
medium and large systems where it is often the only feasible alternative.

Several different numerical approaches can be found in the literature
to calculate $\alpha$ at imaginary frequencies within TDDFT.  For
example, linear response theory can be employed to calculate the
density-density response function $\chi$, from which $\alpha$ directly
follows:
\begin{equation}
  \alpha_{ij}(\omega)= \mint{r} \!\! \mint{r'} r_i \;
  \chi({\bm r}, {\bm r}', \omega) \; r_j \,.
\end{equation}
This approach is quite common and has been used for the calculation of
the $C_6$ coefficients of molecules and
clusters~\cite{vanGisbergen-1995}. Alternatively, one can work in real
time: By propagating the Kohn-Sham equations in real time, it is
immediate to obtain $\alpha(t)$. A Laplace transformation of this
quantity yields $\alpha(\I u)$ and therefore the Hamaker constant
$C_6$. Recently, some of us have shown~\cite{marques-vdw-07} how this
procedure can effectively provide $C_6$ coefficients of large
molecules.  A slightly different approach is the polarization
propagation technique, whose extension to imaginary frequencies has
been used to compute $C_6$ coefficients~\cite{norman01}. For very
large systems, Banerjee and Harbola have proposed the use of orbital
free TDDFT, providing satisfactory results for large sodium
clusters~\cite{banerjee00}.

In this article, we use an alternative scheme based on the solution of
a Sternheimer equation~\cite{sternheimer-1951}. It avoids the use of empty states, 
therefore providing a quite good scaling ($N^2$) with the number of atoms.  The real-time
propagation technique mentioned above has a similar scaling, but we
find the prefactor of the Sternheimer approach to be smaller. This
method has already been used for the calculation of many response
properties, like atomic vibrations, electron-phonon coupling, magnetic
response, etc.~\cite{baroni-2001} In the domain of optical response, it
has been mainly used for static response, although a few calculations
at finite (real) frequency have appeared~\cite{senatore-1987}.

We have implemented the Sternheimer equation at imaginary frequency in
the real-space code {\tt octopus}~\cite{octopus}. The details are
explained in Ref.~~\onlinecite{andrade07} -- although in that case the
equation is solved for real frequencies. A generalization to imaginary frequencies
is straightforward.  The efficiency of the Sternheimer approach is illustrated by the size of
the clusters studied in this Article: up to Si$_{172}$H$_{120}$, i.e.,
$\sim$300 atoms, computed with relatively modest computer systems.
Note that, for these systems, the time required for the evaluation of the van
der Waals coefficients is of the same order as the time required for
the ground-state calculation.

\subsection{Dielectric constant}

The electronic band structure of bulk semiconductors and insulators,
which is the starting point to obtain the dielectric functions, can
nowadays be accurately computed with ab initio
methods~\cite{hedin80}. Much work has been done in the past years to
determine which approximations allow a proper description of
electron-electron and electron-hole interactions, which is essential
to obtain optical functions (at real frequencies) in agreement with
experimental data~\cite{onida-2002}.

The inverse microscopic dielectric function $\epsilon^{-1}$ of a
periodic system is related to the response function $\chi$:
\begin{equation}
\epsilon^{-1}
\left(\textbf{q},\textbf{G},\textbf{G}',\omega \right)
=\delta_{\textbf{G},\textbf{G}'}+v \left( \textbf{q},\textbf{G}
\right) \chi \left( \textbf{q},\textbf{G},
\textbf{G}',\omega \right)\,,
\end{equation}
where $\textbf{q}$ is a vector in the Brillouin zone, $\textbf{G}$ is
a reciprocal lattice vector, and $v$ is the bare Coulomb
interaction. The response function $\chi$ obeys the matrix equation
\begin{eqnarray}
     \chi = \chi_{0} + \chi_{0}\,\left( v+f_{\textrm xc}
     \right)\,\chi , \label{eq:1}
\end{eqnarray}
$\chi_{0}$ being the independent-particle Kohn-Sham response function,
and $f_{\textrm xc}$ the so-called xc kernel.  The macroscopic
dielectric function $\epsilon_{\textrm{M}}$ can be readily obtained
from the microscopic $\epsilon$:
\begin{equation}
    \label{eq:epsmac} \epsilon_{M} \left( \omega
    \right)=\lim_{\textbf{q}\to 0}\frac{1}{\epsilon^{-1} \left(
    \textbf{q},\textbf{G}=0,\textbf{G}'=0,\omega \right)} .
\end{equation} 
The simplest approximation that yields the dielectric function
consists in applying Fermi's golden rule. In this approximation, the
optical spectrum is calculated as a sum of independent transitions
between Kohn-Sham (KS) or quasiparticle states.  This poor man's
approach is known to exhibit severe shortcomings compared to
experiments~\cite{delsole93}.  The next step is the so-called random
phase approximation (RPA), that includes the effects due to the
variation of the Hartree potential upon excitation, while $f_{\textrm
xc}$ is set to zero.  Unfortunately, the RPA does not lead to any
significant improvement for most solids, especially if there are no
particularly pronounced polarizable inhomogeneities in the charge
density. Replacing the KS energies with the quasiparticle energies
does not solve the problem: the peak positions are usually
overcorrected and the oscillator strength is not modified.

It is the neglect of variations of the xc potential, which include the
effect of the electron-hole Coulomb interaction, that is responsible
for an overall disagreement in the absorption strength -- in
particular for the failure to reproduce continuum and bound excitons.
Unfortunately, the adiabatic local density approximation (TDLDA) for
the xc-kernel in the case of solids is not sufficient to yield good dielectric
functions.  The reason for this failure can be traced back to the
short-range nature of the TDLDA $f_{\textrm{xc}}$, while the ``exact''
$f_{\textrm{xc}}$ is expected to be long-ranged~\cite{onida-2002},
decaying in momentum space as $1/q^2$.

A class of kernels that was shown to be yield good results is those
derived from the Bethe-Salpeter equation (BSE)~\cite{onida-2002,
hanke78, botti-review07}, used together with the quasiparticle
bandstructure. A parameter-free expression, the ``nanoquanta kernel'',
was obtained in several different ways (for a detailed discussion we
refer the reader to Ref.~~\onlinecite{botti-review07}, and references
therein). Although involving a potentially reduced computational
effort with respect to the BSE, these calculations are still
significantly more cumbersome than those within the RPA or the
TDLDA. To keep the computational cost as low as possible, in many
cases it is enough to use simplified versions of this kernel. It was
shown that the nanoquanta kernel has the asymptotic form of a
long-range contribution (LRC)~\cite{reining02, botti04}
\begin{equation}
  f^{\textrm{static}}_{\textrm{xc}}(\textbf{q}) =
  -\frac{\alpha^{\textrm{static}}}{q^2},
\label{eq:fxc-LRC}
\end{equation}
where $\alpha^{\rm static}$ is a material dependent parameter, that
can be related to the dielectric constant. This long-range
contribution alone is sufficient to simulate the strong continuum
exciton effect in the absorption spectrum and in the refraction index
of several simple semiconductors, like bulk silicon or GaAs, provided
that quasiparticle energies are used as a starting point. A dynamical
extension of this LRC model~\cite{botti05} of the form
\begin{equation}
  f^{\textrm{dyn}}_{\textrm{xc}}(\textbf{q}) = -\frac{\alpha + \beta \omega^2}{q^2}
  \label{eq:fxc-LRC-dyn}
\end{equation}
leads to remarkable improvements for optical spectra of large gap
systems with respect to calculations where the kernel is imposed to be
static. Moreover, the dynamical approach was proved to be valid also
for energies in the range of plasmons and for the determination of
dielectric constants.  Note that the parameters of both the static and
the dynamical model can be related to physical quantities, like the
experimental dielectric constant and the plasmon frequency.

In this work, we calculated the dielectric functions at imaginary
frequency using the computer code DP~\cite{dp}, an ab initio linear
response, plane wave, TDDFT code. Despite the enormous amount of
studies concerning the accuracy of different approximations for the xc
kernel for solids, it is not a priori clear which approximation is
more suitable when one wants to work at imaginary frequencies.  In
order to clarify this aspect, we tested several approximations for the
xc kernel.

\section{Calculations}
\label{sec:calculations}

\subsection{van der Waals  interactions between silicon clusters}

\begin{figure}[t]
  \centerline{\includegraphics[width=1.0\columnwidth]{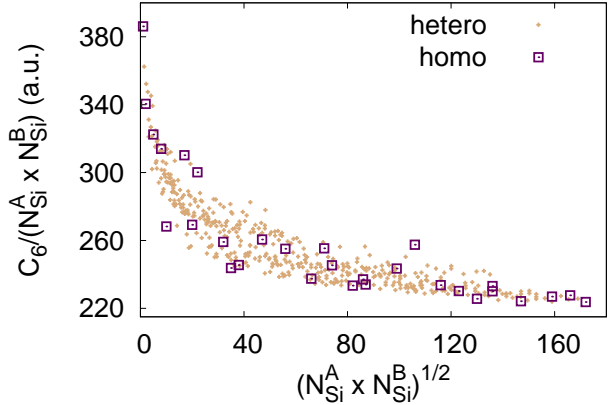}}
  \caption{ \label{fig:vdw_all_c6} (Color online) $C_6^{AB}$ Hamaker constants as a
  function of the square root of the product of silicon atoms in
  cluster A and B. As $C_6^{AB}$ scales basically with the product
  between the number of (silicon) atoms in A and B, we plot the
  Hamaker constants divided by this number. Values of $C_6^{AB}$ when
  A and B are the same cluster are plotted as red (dark gray) squares, otherwise
  they are plotted as orange (light gray) dots.}
\end{figure}

\begin{figure}[t]
  \centerline{\includegraphics[width=1.0\columnwidth]{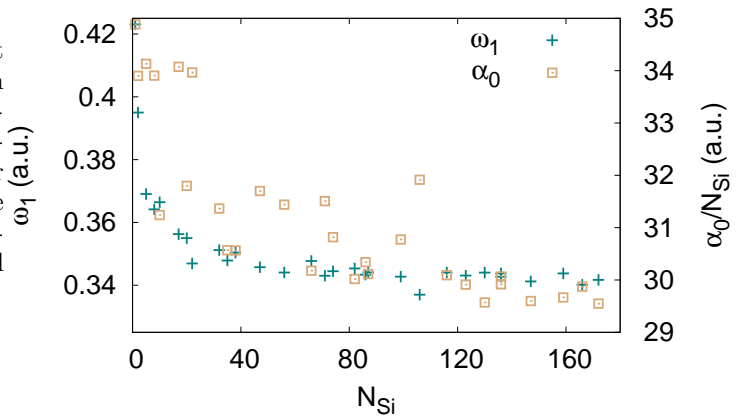}}
  \caption{ \label{fig:vdw_wI} (Color online) London effective frequency $\omega_1$ --
    blue (dark gray) crosses, and static polarizabilities per atom -- red
    (light gray) squares, for the silicon clusters under study, as a function of
    the number of silicon atoms.  }
\end{figure}

We start our discussion by the calculation of C$_6$ between the
silicon clusters. The clusters were cut from bulk silicon, and then
saturated with hydrogens along the tetrahedral direction of the
surface atoms.  The geometries were optimized with the computer code
siesta~\cite{siesta}, employing norm-conserving pseudopotentials,
a double $\zeta$ with polarization
basis set, and the PBE parametrization~\cite{pbe96} for the
xc potential.

From the optimized geometries, we then obtained the electric
polarizability within TDDFT using the Sternheimer equation, as
implemented in the computer code {\tt octopus}~\cite{octopus}. The
electron-ion interaction was described through norm-conserving
pseudopotentials~\cite{troullier91} and the local density
approximation (LDA)~\cite{perdew-zunger81} was employed in the
adiabatic approximation for the xc potential.  It is known that the
LDA provides reliable results for semiconducting
clusters~\cite{vasiliev-2000}; Furthermore, from previous experience
with optical spectra calculations~\cite{marques-2001}, we know that
these results will not change significantly with the use of the more
sophisticated GGAs. The equations, in this code, are represented in a real-space regular grid, 
whose spacing is chosen to be 0.275 A. The simulation box is constructed 
by joining spheres of radius 4.5 A, centered around each atom. The integrals in Eqs.~\eqref{eq:c6}
and \eqref{eq:c3} were performed with a Gauss-Legendre quadrature
using 6 frequency values. With these parameters, we estimate the
accuracy of our numerical calculations to better than 5\%.

In Fig.~\ref{fig:vdw_all_c6} we show our results for the C$_6$ Hamaker
constant between silicon clusters. As the value of C$_6$ scales with
the product of the atoms in the cluster A ($N_{\textrm{Si}}^{A}$) and
in cluster B ($N_{\textrm{Si}}^{B}$), we divided the Hamaker constant
by $N_{\textrm{Si}}^{A} N_{\textrm{Si}}^{B}$ to eliminate this
dependence. We show both constants between two identical clusters
(homo-molecular -- as red (dark gray) squares, also presented in
Table~\ref{tab:alphaC6}) and between different clusters
(hetero-molecular -- orange (light gray) dots). We see that the largest C$_6$ per
atom squared comes from the interaction between two SiH$_4$ clusters,
and then the values decrease rapidly until slightly above 220 a.u.\!,
where it saturates.  The few clusters that fall far from the line are
the most asymmetric, for which a description in terms of the average
of the dipole polarizability tensor is not necessarily as good.

The polarizability at imaginary frequencies can be modelled in the
London approximation by introducing two adjustable parameters: the
static polarizability $\alpha(0)$ and one effective frequency
$\omega_1$:
\begin{equation}
  \alpha({\rm i}u) = \frac{\alpha(0)}{1+(u/\omega_1)^2}\,.
  \label{eq:London-model}
\end{equation}
If we insert (\ref{eq:London-model}) in (\ref{eq:c6}) we obtain a
simplified expression for the homo-molecular Hamaker constant in terms
of these parameters:
\begin{equation}
\label{eq:w1}
  C_6 = \frac{3\omega_1}{4}\alpha^2(0)\,.
\end{equation}
As we have calculated both $C_6$ and $\alpha(0)$ within TDLDA, it is
easy to extract $\omega_1$ from Eq.~(\ref{eq:w1}). The resulting
effective frequencies are plotted in Fig.~\ref{fig:vdw_wI}, together
with the calculated static polarizabilities per number of Si atoms.
We can observe that $\omega_1$ decreases with the number of Si atoms,
but the dependence on the size of the cluster is rather weak, except
for the singular case of the smallest aggregates.

\begin{figure}[t]
  \centerline{\includegraphics[width=1.0\columnwidth]{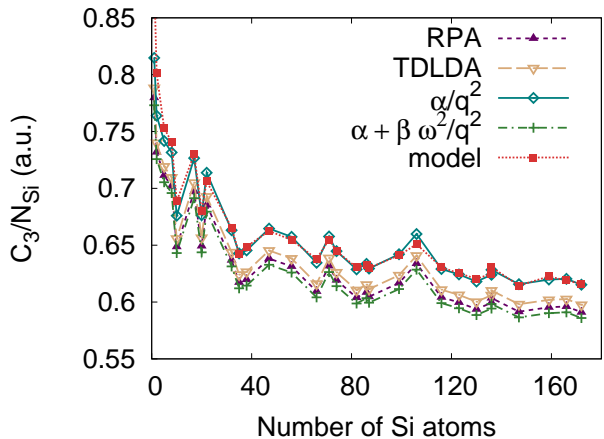}}
  \caption{ \label{fig:c3_Si} (Color online) Van der Waals C$_3$ coefficients between
  silicon nanoclusters and a silicon surface. The C$_3$ coefficients
  were divided by the number of silicon atoms in the cluster. The
  different curves were calculated using different approximations for
  the dielectric constant of the bulk crystal at imaginary frequencies
  (see the text for details).}
\end{figure}

\begin{figure}[t]
  \centerline{\includegraphics[width=1.0\columnwidth]{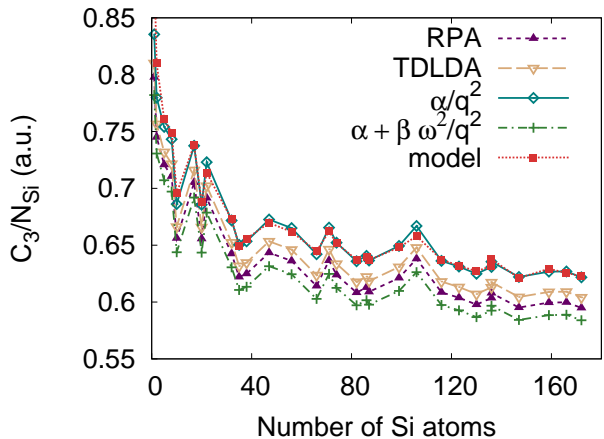}}
  \caption{ \label{fig:c3_SiC} (Color online) Van der Waals C$_3$ coefficient between
  silicon nanoclusters and a silicon carbide surface. The meaning of
  the curves is as in Fig.~\ref{fig:c3_Si}.}
\end{figure}

\begin{figure}[t]
  \centerline{\includegraphics[width=1.0\columnwidth]{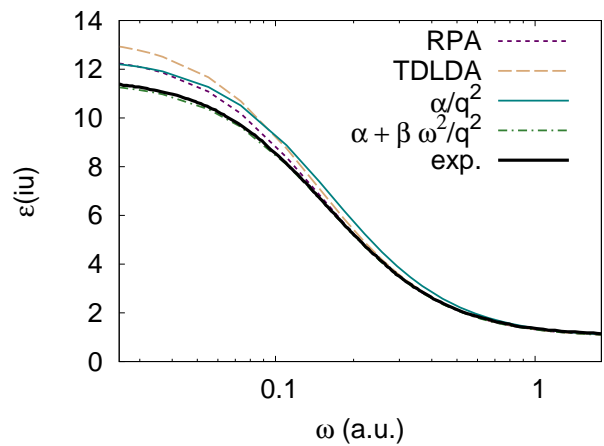}}
  \caption{(Color online) Dielectric function of Si along the imaginary axis as a
  function of the frequency in a logarithmic scale. Results obtained
  with different approximations for the xc kernel are compared to the
  curve extracted from experimental data by using Eq.~(\ref{eq:KK}.)
  \label{fig:epsilon_Si}}
\end{figure}

\begin{figure}[t]
  \centerline{\includegraphics[width=1.0\columnwidth]{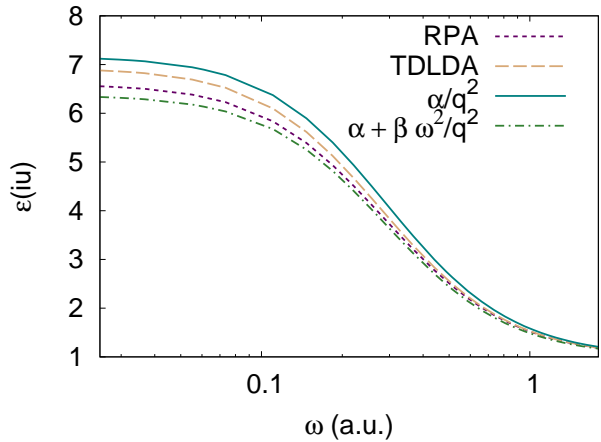}}
  \caption{(Color online) Dielectric function of SiC along the imaginary axis as a
  function of the frequency in a logarithmic scale. We compare results
  obtained within different approximations for the xc kernels.
    \label{fig:epsilon_SiC}~\cite{caride05}}
\end{figure}

\subsection{van der Waals interactions between silicon clusters and
  dielectric surfaces}

Next, we consider the case of a silicon cluster in proximity of a
surface of Si or SiC in the zincblende phase.  In this case we want
to calculate $C_3$ coefficients, that are determined both by the
dynamical polarizability of the cluster and the dielectric function of
the bulk crystal.

The ground state calculations for the bulk crystals were performed
using the plane-wave code ABINIT~\cite{abinit} with norm-conserving
Hamann pseudopotentials~\cite{hamman} for Si and C. We used a cutoff
energy for the plane wave basis of 12.5\,Ha for Si and 30\,Ha for SiC.
The unit cell was relaxed within the LDA approximation, yielding
lattice parameters with an error smaller than 3\%.  The Kohn-Sham energies
and wavefunctions yielded by ground state calculations were employed to 
calculate dielectric functions at imaginary frequencies using the code
DP~\cite{dp}.  For the response calculations a shifted $\textbf{k}$-point grid of 256 points
was used both for Si and SiC. More detailed information on the numerics and convergence
issues can be found in Ref.~~\onlinecite{botti04}.

As discussed before, previous tests on the effect of different
approximations for the xc potential demonstrate that the dynamical
polarizability of the hydrogenated Si clusters is accurately described
within the TDLDA, and therefore the $C_6$ coefficients are not going
to change by more than 5\% by using different approximations.  We
decide thus to focus on the effect of different models for the
xc-kernel in the calculation of C$_3$.

In Figs.~\ref{fig:c3_Si} and \ref{fig:c3_SiC} we can compare $C_3$
coefficients for Si clusters on a Si surface and Si clusters on a SiC
surface, respectively. We present results obtained within the RPA
(violet upright triangles), the TDLDA (beige inverted triangles),
using the static LRC kernel (blue diamonds) and the dynamical LRC
kernel (green crosses). Note that, in the case of the RPA and the
TDLDA calculations, we used the Kohn-Sham band structure to build
$\chi_0$, while the GW quasiparticle states are used when the static
or dynamical LRC kernels are employed (see Refs.~~\onlinecite{botti04}
and ~\onlinecite{botti05} for details).  In Si and SiC the GW
corrections to the bandstructures are essentially equivalent to a
rigid shift of the conduction states, thus replacing KS energies with
quasiparticle energies leads to a rigid shift of the absorption
spectrum towards higher energies. We also plotted in
Figs.~\ref{fig:c3_Si} and \ref{fig:c3_SiC} the values of $C_3$
obtained using simple models (red squares) for both the dynamical
polarizability of the cluster and the dielectric function of the
crystal at imaginary frequencies. We will discuss these analytical
models and the quality of their results in Section~\ref{sec:models}.
The peaks of $C_3$ as a function of the number of Si atoms 
occur for higher polarizable clusters. The oscillations of the polarizabilities in turn
exactly correlate with the binding energy, with largest polarizabilities corresponding
to the most stable clusters.

For the interaction of Si clusters on either a Si or a SiC surface,
all the approximations used for the xc kernel give curves with very
similar trends and a dispersion of the values which is smaller than
10\%. This finding reflects the fact that the dielectric function at
imaginary frequency is a very smooth and well behaved curve, and
therefore fairly simple to reproduce; it starts at the value of the
static dielectric constant at $\I u = 0$, and then decreases
monotonically to its asymptotic limit of one (see
Figs.~\ref{fig:epsilon_Si} and
\ref{fig:epsilon_SiC}).

Dielectric constants in the imaginary frequency axis can be obtained
experimentally by performing a Kramers-Kronig transformation of the
values obtained in the real axis,
\begin{equation}
 \epsilon_{\textrm{M}}(\textrm{i} u) = 1 + \frac{2}{\pi} \int_0^{\infty}
 \!\!\!{\textrm{d}}\omega \; \frac{\omega
 \Im[\epsilon_{\textrm{M}}(\omega)]}{\omega^2 + u^2}
\label{eq:KK}
\end{equation}
as long as the experimental absorption spectra has been measured on a
large enough spectral range. We include in Fig.~\ref{fig:epsilon_Si}
the experimental curve for Si~\cite{caride05}, for comparison.  The
curve calculated with the dynamical LRC approximation is exactly
superposed to the experimental curve.  In fact, this is the only
approximation that yields a good dielectric constant, which fixes the
interception with the y axis, and an overall good shape of the
absorption spectrum over a large spectral range~\cite{botti05}.

It is interesting to notice that the static LRC results are worse than
even the RPA curve, being overestimated over the whole frequency
range.  In the RPA, there is a ``fortuitous'' compensation of errors
(the error due to the too high dielectric constant is balanced by the
shift of the spectral weight to lower energies due to the DFT-LDA underestimation
of the absorption edge).  The TDLDA curve is
the one that lays further from the dynamical LRC solution at lower
frequencies due to the even higher dielectric constant, but it greatly
improves for $u \ge 0.1$, thanks to the same compensation of errors
already observed for the RPA calculation.

The same conclusions can be obtained for SiC (see
Fig.~\ref{fig:epsilon_SiC}). In this case, we do not have access to
experimental results, but it is reasonable to expect that the
dynamical LRC approximation will yield the most accurate result
overall. The static LRC results again shows a consistent
overestimation of the dielectric function, while the the RPA curve is
the closest to the dynamical LRC result.

In the light of this analysis, one can interpret the results for the
$C_3$ coefficients. First of all, the dynamical LRC kernel is expected
to work very well. Outside the limits of validity of the dynamical LRC
model (large gap insulators, strongly bound excitons) only a
calculation for the bulk crystal based on the solution of the BSE (or,
equivalently, based on the fully ab initio Bethe-Salpeter derived
kernel) can guarantee the quality of the $C_3$ coefficients.  This
implies necessarily larger computational costs. Perhaps surprisingly,
the RPA and TDLDA appear to be good approximations to evaluate $C_3$,
despite their well known deficiencies in the calculation of optical
absorption spectra. This is not necessarily true for every system, but
it is probably true provided that the calculated dielectric function at
zero frequency is larger than the experimental one.  In this case, in fact,
we can expect the (at least partial) cancellation of error between the
too high starting point of the curve $\epsilon(\I u)$ and its too fast decay
to one as a consequence of the shift of the spectral weight to lower energies.
One should be very careful not to use the
static LRC approximation for the kernel, despite the fact that it
gives an absorption spectrum in overall agreement with the experiment.
This is due to the fact that the van der Waals coefficients are very
sensitive to the value of the dielectric constant.

\section{Models}
\label{sec:models}

Our proposed ab initio techniques are quite efficient and allow the
calculation of van der Waals coefficients for systems with $\sim$300
atoms, even in relatively modest computer systems. Moreover, the
crystals of Si and SiC considered here contain only two atoms per unit
cell and allow very fast calculations.  Nevertheless, a full ab initio
study of dispersion interactions of large nanostructures/biological
molecules on complex surfaces can become a computationally demanding
task. Therefore, it is desirable to design accurate model van der
Waals potentials, based on ab-initio calculations, to be used for
these calculations, where a number of atoms of the order of 1000 and
even larger can be easily attained. Two problems need to be addressed:
i) how to model the dynamic polarizability at imaginary frequency of
the nanoobject, ii) how to model the dielectric function at imaginary
frequency of the solid.

\begin{table*}[t]
  \setlength{\tabcolsep}{0.25truecm}
  \caption{
    \label{tab:alphaC6}
    Values for the static polarizability $\alpha(0)$, and for the Hamaker C$_6$
    coefficient. We show both the calculated values with (TD)DFT, as well as the
    values obtained using a bond polarization model (BPM) and effective medium
    theory (EMT), and the respective percent errors.
  }
  \centering
  \begin{tabular}{lrrrrrrrr}
    & \multicolumn{3}{c}{$\alpha(0)\times 10^{-2}$} &
    \multicolumn{3}{c}{C$_6\times 10^{-4}$} & \multicolumn{2}{c}{$\Delta$C$_6$/C$_6$ [\%] }   \\
    & \multicolumn{1}{c}{DFT} & \multicolumn{1}{c}{BPM} & \multicolumn{1}{c}{EMT} &
    \multicolumn{1}{c}{DFT} & \multicolumn{1}{c}{BPM}
    &\multicolumn{1}{c}{EMT}   &  \multicolumn{1}{c}{BPM} & \multicolumn{1}{c}{EMT} \\
    \cline{2-4} \cline{5-7} \cline{8-9}\\[-2mm]
    SiH$_4$             & 0.349 & 0.422 & 0.337 & 0.0386 & 0.0458 & 0.0340 & 19 & -11 \\
    Si$_{2  }$H$_{6  }$ & 0.678 & 0.768 & 0.632 & 0.136 & 0.151 & 0.119 & 11 & -9\\
    Si$_{5  }$H$_{12 }$ & 1.71  & 1.80  & 1.52 & 0.806 & 0.836 & 0.686 & 4 & -15 \\
    Si$_{8  }$H$_{18 }$ & 2.71  & 2.84  & 2.40 & 2.01 & 2.07 & 1.72 & 3& -14\\
    Si$_{10 }$H$_{16 }$ & 3.12  & 3.30  & 2.86 & 2.68 & 2.80 & 2.45 & 4 & -9\\
    Si$_{17 }$H$_{36 }$ & 5.79  & 5.95  & 5.05 & 8.97 & 9.09 & 7.62 & 1 & -15\\
    Si$_{20 }$H$_{30 }$ & 6.36  & 6.52  & 5.68 & 10.8 & 10.9 & 9.64 & 1 & -10\\
    Si$_{22 }$H$_{40 }$ & 7.47  & 7.44  & 6.40 & 14.5 & 14.2 & 12.23 & -2 & -16 \\
    Si$_{32 }$H$_{42 }$ & 10.0  & 10.2  & 8.96 & 26.5 & 26.7 & 24.0 &  1 & -10\\
    Si$_{35 }$H$_{36 }$ & 10.7  & 10.8  & 9.59 & 29.9 & 29.8 & 27.49 & -0.3 & -8 \\
    Si$_{38 }$H$_{42 }$ & 11.6  & 11.8  & 10.5 & 35.5 & 35.9 & 32.8 & 1 & -8\\
    Si$_{47 }$H$_{60 }$ & 14.9  & 14.9  & 13.1 & 57.6 & 57.2 & 51.5 & -1 & -11\\
    Si$_{56 }$H$_{66 }$ & 17.6  & 17.6  & 15.5 & 80.0 & 79.3 & 72.0 & -1 & -10\\
    Si$_{66 }$H$_{64 }$ & 19.9  & 20.2  & 18.0 & 103  & 105  & 96.8 & 2 & -6\\
    Si$_{71 }$H$_{84 }$ & 22.4  & 22.3  & 19.7 & 128  & 128  & 116 & 0 & -10\\
    Si$_{74 }$H$_{78 }$ & 22.8  & 22.9  & 20.3 & 134  & 134  & 123 & 0 & -8\\
    Si$_{82 }$H$_{72 }$ & 24.6  & 24.8  & 22.2 & 156  & 158  & 147 & 1 & -6\\
    Si$_{86 }$H$_{78 }$ & 26.1  & 26.1  & 23.4 & 175  & 175  & 163 & 0 & -7\\
    Si$_{87 }$H$_{76 }$ & 26.2  & 26.3  & 23.6 & 177  & 177  & 166 & 0 & -6\\
    Si$_{99 }$H$_{100}$ & 30.5  & 30.4  & 27.1 & 238  & 238  & 219 & 0 & -8\\
    Si$_{106}$H$_{120}$ & 33.8  & 33.1  & 29.3 & 289  & 281  & 256 & -3 & -11\\
    Si$_{116}$H$_{102}$ & 34.9  & 35.1  & 31.4 & 314  & 316  & 295 & 1 & -6\\
    Si$_{123}$H$_{100}$ & 36.8  & 36.9  & 33.2 & 348  & 349  & 328 & 0.3 & -6\\
    Si$_{130}$H$_{98 }$ & 38.4  & 38.6  & 34.9 & 381  & 384  & 363 & 1 & -5\\
    Si$_{136}$H$_{110}$ & 40.7  & 40.7  & 36.6 & 425  & 426  & 401 & 0.2 & -6\\
    Si$_{136}$H$_{120}$ & 40.9  & 41.1  & 36.9 & 431  & 434  & 406 & 1 & -6\\
    Si$_{147}$H$_{100}$ & 43.5  & 43.3  & 39.2 & 484  & 481  & 459 & -1 & -5\\
    Si$_{159}$H$_{124}$ & 47.2  & 47.4  & 42.7 & 573  & 578  & 546 & 1 & -5\\
    Si$_{166}$H$_{122}$ & 49.6  & 49.2  & 44.5 & 627  & 623  & 590 & -1 & -6\\
    Si$_{172}$H$_{120}$ & 50.8  & 50.8  & 45.9 & 661  & 662  & 630 & 0.2 & -5\\
  \end{tabular}
\end{table*}

\begin{figure}[t]
  \centerline{\includegraphics[width=1.0\columnwidth]{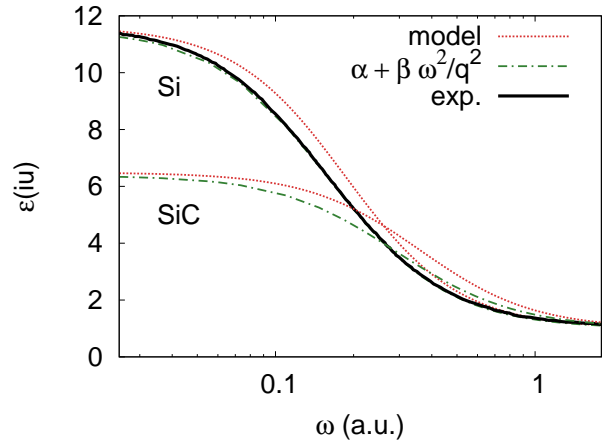}}
  \caption{(Color online) Dielectric function of Si and SiC along the imaginary axis:
  the model function is compared with the curve obtained using the LRC
  dynamical approximation for the xc kernel, which gives the best
  agreement with the experiment.  \label{fig:epsilon_Si_SiC}}
\end{figure}

\subsection{Model for the dielectric function}

Let us start by the modelling of the semiconductor or insulator
surface. We have seen that the simplest expression for the
longitudinal dielectric function can be derived by applying
Fermi's golden rule and assuming the case of a (nearly) homogeneous
material:
\begin{equation}
  \label{eq:long_dielectric_function} \epsilon ( \textbf{q}, \omega) =
  1 + \frac{8 \pi}{q^2} \frac{1}{V} \sum_{\beta \gamma} \frac{\left|
  \langle \psi_{\gamma} \left| e^{\textrm{i} \textbf{q} \cdot
  \textbf{r} } \right| \psi_{\beta} \rangle \right|^2}
  {\omega_{\gamma} - \omega_{\beta} - \omega - \textrm{i} \eta }
  \left[ f(\omega_{\beta}) - f(\omega_{\gamma}) \right] \,.
\end{equation}
We have observed in Section~\ref{sec:calculations} that the RPA
approximation gives a rather good dielectric function at imaginary
frequency thanks to a compensation of errors. It is thus reasonable to
start from this simple approximation to design an analytical model.  A
crude approximation consists in replacing the energy differences in
Eq.~(\ref{eq:long_dielectric_function}) with some average excitation
energy $\omega_{\textrm{av}}$. By exploiting the sum rules and the
asymptotic behavior of $\epsilon_{\textrm{M}} (\omega)$ one gets in
the optical limit ($q \rightarrow 0$)~\cite{grosso-pastori00}:
\begin{equation}
  \label{eq:1-oscillator-model-real} \epsilon_{\textrm{M}} (\omega) = 1 -
  \frac{\omega_\textrm{p}^2}{\omega^2 - \omega_{\textrm{av}}^2  +
  \textrm{i} \eta \omega} \qquad \textrm{with} \quad \eta \rightarrow 0^+
  \,.
\end{equation}
Equation~(\ref{eq:1-oscillator-model-real}) has the same form as the Lorentz
dielectric function of bound charged carriers with frequency
$\omega_{\textrm{av}}$. This model predicts for the static dielectric
constant
\begin{equation}
  \label{eq:static-dielectric-constant} \epsilon_{\textrm{M}} (0) = 1 +
  \frac{\omega_{\textrm{p}}^2}{\omega_{\textrm{av}}^2} \,.
\end{equation}
This equation can be used together with the value of the plasma
frequency $\omega_{\textrm{p}} = 4 \pi N_{\textrm{el}}/V$ to estimate
$\omega_{\textrm{av}}$. In this approximation one only
needs to fix two parameters that are easily accessible from
experiments, the static dielectric constant and the volume of the unit
cell.
At imaginary frequencies an analogous expression can be derived
as a function of the same parameters:
\begin{equation}
  \label{eq:1-oscillator-model-imag} \epsilon_{\textrm{M}} (\textrm{i} u) = 1 +
  \frac{\omega_\textrm{p}^2}{\omega_{\textrm{av}}^2 + u^2} \,.
\end{equation}
We remind that $\epsilon_{\textrm{M}} (\textrm{i} u)$ is a real
function.  The resulting model dielectric function is plotted in
Fig.~\ref{fig:epsilon_Si_SiC} for both Si and SiC, together with the
best theoretical curve (the one obtained using the dynamical LRC model
for the xc kernel) and the experimental curve for Si. The model
function is substantially worse than the calculated curves, even with
respect to the RPA calculations.

\subsection{Models for the cluster dynamical polarizabilities}

For the atomic polarizabilities at imaginary frequency $\alpha
(\textrm{i}u)$ it is convenient to use to already mentioned London
approximation, Eq.~(\ref{eq:London-model}).  In that case, two
parameters for \textit{every} cluster have to be fixed: its static
polarizability $\alpha(0)$ and the energy of an effective frequency
$\omega_1$. Of course, one does not want to define a different set of
parameters for every cluster, but only two parameters for all
possible clusters of a fixed species.  In the case of a larger
molecule or a cluster, the static polarizability can be estimated by
making use of the bond polarization model (BPM)~\cite{wolkenstein41,cardona82} as suggested by
Jiemchooroj \textit{et al.}~\cite{jiemchooroj04} In this model, the
total polarizability is obtained summing over the contributions from
the individual polarizable entities: the covalent bonds.  In our case
there are only two kinds of bonds, Si-Si and Si-H, and we can write
the static polarizability as
\begin{equation}
  \label{eq:additivity} \alpha_{i} (0) = n_{i}^{\textrm{Si-Si}}
  \alpha_{\textrm{Si-Si}} + n_{i}^\textrm{Si-H} \alpha_{\textrm{Si-H}}
  \,,
\end{equation}
where $n_{i}^{\textrm{Si-Si}}$ and $n_{i}^\textrm{Si-H}$ are the
number of Si-Si and Si-H bonds, respectively, of the cluster $i$. Here
we have indicated with $\alpha_{\textrm{Si-Si}}$ and
$\alpha_{\textrm{Si-H}}$ the contributions to the polarizability due
to the Si-Si and the Si-H bonds, respectively.  Upon substitution in
the integral (\ref{eq:c6}), the London model gives for a
homo-molecular Hamaker constant the expression (\ref{eq:w1}), and for
the hetero-molecular Hamaker constant:
\begin{equation}
  \label{eq:hetero-C6-London}
  C_6^{ij} = 3 \frac{\bar\omega_1^i \bar \omega_1^j}{2 ( \bar\omega_1^i + \bar \omega_1^j)}
  \alpha_{i} (0) \alpha_{j} (0) \,,
\end{equation}
where (\ref{eq:additivity}) gives the values of $\alpha_i(0)$.
If we perform a set of ab-initio calculations of $C_6^{ii}$ and
$\alpha_i(0)$ for small-medium size clusters, we can extract
$\omega_1^i$ from Eq.~(\ref{eq:w1}) and the parameters
$\alpha_{\textrm{Si-Si}}$, $\alpha_{\textrm{Si-H}}$ by fitting the
theoretical curve for the static polarizability with
Eq.~(\ref{eq:additivity}).

The small dispersion of values for $\omega_1^i$ in
Fig.~\ref{fig:vdw_wI} suggests that it is possible to determine a
single average frequency $\bar\omega_1=$0.343~Ha for all clusters.
For the small nanocrystals with less than 10 Si atoms this
approximation is not very precise, but we are interested in getting
information on the interaction of larger systems, that cannot be
easily studied by ab-initio techniques.  The parameters
$\alpha_{\textrm{Si-Si}}$ and $\alpha_{\textrm{Si-H}}$ are fixed by
fitting Eq.~(\ref{eq:additivity}) to the curve for the static
polarizability calculated within the TDLDA (see
Fig.~\ref{fig:vdw_wI}). The outcome are the values
$\alpha_{\textrm{Si-Si}}=$13.41~a.u.  and
$\alpha_{\textrm{Si-H}}=$10.56~a.u. In Table~\ref{tab:alphaC6} we can
verify the excellent agreement ($\left| \Delta \alpha(0)
\right| \leq 1 \%$ for all the clusters, excluding the
smallest ones) between the static polarizabilities calculated using
TDLDA and the additivity model for the nanocrystals.
The same agreement is conserved for the estimated values of $C_6$: For
the big clusters, the difference with the calculated values is
remarkably small (see Table~\ref{tab:alphaC6}).

A different approach to model the dynamic polarizability of a
nanocrystal is to start from the dielectric function of the
corresponding bulk crystal and apply the effective medium theory
(EMT)~\cite{emt}. This classical approach is based on the solution of
Maxwell's equations with the assumption that the dielectric response
of each constituent of the system is the one of the corresponding
bulk. This assumption is better justified when the size of the
composing objects is large. In fact, EMT completely neglects the
microscopic scale details, such as atoms and bonds. In this respect it
is complementary to the additive procedure. However, it handles
correctly the boundary conditions for the Maxwell's equations at the
interfaces, which give very important contributions to the dielectric
response through the crystal local field effects.  Our clusters can be
considered as a sphere of Si in vacuum with a filling factor $f$
that goes to zero.  The Maxwell-Garnett expression~\cite{emt} yields
in this specific case:
\begin{equation}
 \Im{\left\{\alpha \left( \omega \right) \right\}} = \frac{-9 V_{\textrm{s}}}{4 \pi} 
\Im{ \left\{ \frac{1}{\epsilon_{\textrm{M}}^{\textrm{Si}} \left( \omega \right) +2} \right\}}  \,,
\end{equation}
where $\epsilon_{\textrm{M}}^{\textrm{Si}}$ is the complex dielectric
function of bulk silicon, and where $V_{\textrm{s}}$ is the volume of
the spherical cluster.  By applying the analogous of the Laplace
transformation (\ref{eq:KK}) to the dynamical polarizability at real
frequencies and using once again the single-oscillator model
(\ref{eq:1-oscillator-model-real}) for the dielectric function of bulk
Si, we obtain for the dynamical polarizability at imaginary
frequencies:
\begin{equation}
 \alpha \left( \I u \right) = \frac{V_{\textrm{s}}}{4 \pi} \frac{\omega_{\textrm{p}}^2}{u^2 + 
\omega_{\textrm{av}}^2 + \omega_{\textrm{p}}^2/3 } \,,
\end{equation}
This
expression can be rewritten in the same form as the London model by
imposing:
\begin{equation}
\alpha (0) =  \frac{V_{\textrm{s}} }{4 \pi} \frac{\omega_{\textrm{p}}^2}{\omega_{\textrm{av}}^2 + \omega_{\textrm{p}}^2/3} 
\end{equation}
and
\begin{equation}
 \omega_1 = \sqrt{\omega_{\textrm{av}}^2 + \omega_{\textrm{p}}^2/3} \,.
\end{equation}
With respect to the model for the dielectric function of the crystal
we have here an extra parameter to be estimated: the volume of the
spherical cluster. In the limit of a very large cluster, we can assume
that the volume per Si atom is the same as in the bulk crystal. To
obtain better results for small-medium sized clusters it is necessary
to include the contribution to the volume due to the hydrogen atoms at
the surface (considering a Si-H bond distance of about 1.5\AA, we can
assume a volume of about 11.4\,a.u. per hydrogen atom).  The value of
$\omega_1$ is 0.4\,Ha, which is 20\% larger than $\omega_1$ evaluated
in the BPM. The values of $\alpha(0)$, on the other hand, are
systematically smaller than their counterparts in the BPM. As a
result, $C_6$ coefficients are underestimated by about 5\% for the
larger clusters and up to 10-15\% for the smaller ones.

The non-trivial advantage of this second approach is the fact that no
ab-initio calculation needs to be performed to fit $\alpha(0)$ and
$\omega_1$.

\subsection{Modelled $C_3$ coefficients}

Finally, we calculated the $C_3$ coefficients both for Si nanocrystals
on Si surfaces and on SiC surfaces combining the single-oscillator
model for the surface the the bond polarization model for the
nanocrystal.  The curves obtained are plotted in Figs.~\ref{fig:c3_Si}
and \ref{fig:c3_SiC} where they can be compared with the results of
the ab-initio calculations. The model calculations are rigidly shifted
to higher values by about 5\% for Si and 7\% for SiC (when compared
with the dynamical LRC model).  Curiously, they almost overlap with
the results obtained using the LRC xc kernel for the determination of
the dielectric function. This can be understood by inspecting
Fig.~\ref{fig:epsilon_Si_SiC}: the error is likely to be entirely due
to the insufficiently accurate description of the dielectric function
of the bulk material.

\section{Conclusions}
\label{sec:conclusions}

We have demonstrated how the leading terms of the van der Waals forces
acting between nanostructures, and between nanostructures and
non-metallic surfaces, can be accurately and inexpensively computed
from first principles. The key ingredients can be reliably obtained
with state-of-the-art theoretical schemes and computational
procedures. The dynamical polarizabilities of large nanostructures at
imaginary frequencies, for example, can be safely computed with the
Sternheimer reformulation of time-dependent density-functional theory.
Regarding the other key ingredient necessary to obtain the
cluster-surface interaction, the macroscopic dielectric constant, it can
also be computed by making use of TDDFT.  In this case, special care
has to be taken with the choice of the xc kernel. We have found that a
particularly simple and reliable scheme is to make use of the
dynamical ``long-range contribution'' (LRC)
kernel~\cite{botti05}. However, if the bulk to be studied lies outside
the limits of validity of the dynamical LRC model (large gap
insulators, systems with strongly bound excitons), one probably has to
resort to the full solution of the Bethe-Salpeter equation (or,
equivalently, to a TDDFT calculation based on a kernel derived from
the Bethe-Salpeter equation~\cite{botti-review07}).  This necessarily implies
larger computational costs.

We also suggest some simplified models that should supply reasonable
estimates for the van der Waals coefficients, for those cases in which
first principle calculations are out of range. We have found that
modelling the bulk dielectric function can lead to a substantial error
($\sim10$\%). However, using a bond polarization model or the
effective medium theory for the nanostructure dynamical
polarizability, yields very precise results, especially for large
systems. This opens the way to the simulation of the van der Waals
interaction for large nanocrystals.

\acknowledgments

All calculations were performed at the Laborat\'{o}rio de
Computa\c{c}{\~a}o Avan\c{c}ada of the University of Coimbra. The
authors were partially supported by the EC Network of Excellence
NANOQUANTA (NMP4-CT-2004-500198).
S. Botti acknowledges financial support from French ANR (JC05–46741).
A. Castro acknowledges financial
support from the Deutsche Forschungsgemeinschaft within the SFB
658. X. Andrade acknowledge partial support from the EU Programme
Marie Curie Host Fellowship (HPMT-CT-2001-00368). M.A.L. Marques
acknowledges partial support by the Portuguese FCT through the project
PTDC/FIS/73578/2006. X. Andrade and A. Rubio acknowledge financial support from the Spanish
Ministry of Education (Grant No. FIS2007-65702-C02-01) and 
Grupos Consolidados UPV/EHU of the Basque Country Government (2007).


\begin{thebibliography}{99}

\bibitem{vanderwaals} J. Israelachvili, {\it Intermolecular and Surface
    Forces} (Academic Press, San Diego, 1992); J. Mahanty and
  B. W. Ninham, {\em Dispersion Forces} (Academic Press, New York,
  1976).

\bibitem{vanderwaals-historical} J. D. van der Waals, Ph.D. thesis,
  University of Leiden, 1873; translated to English by R. Threlfall
  and J. F. Adair, in {\it Physical Memoirs, Selected and Translated
    from Foreign Sources}, 1 (Physical Society, London, 1890), 333.

\bibitem{tian-2006} Y. Tian, N. Pesika, H. B. Zeng, K. Rosenberg, B.
  X. Zhao, P. McGuiggan, K. Autumn, J. Israelachvili, Proc. Natl.
  Acad. Sci. USA {\bf 103}, 19320 (2006).

\bibitem{binnig-1986} G. Binnig, C. F. Quate and Ch. Gerber,
  Phys. Rev. Lett. {\bf 56}, 930 (1986).

\bibitem{shimizu-2001} F. Shimizu, Phys. Rev. Lett. {\bf 86}, 987
  (2001); V. Druzhinina and M. DeKieviet, Phys. Rev. Lett. {\bf 91},
  193202 (2003).

\bibitem{harber-2003} D. M. Harber, J. M. McGuirk, J. M. Obrecht and
  E. A. Cornell, J. Low. Temp. Phys. {\bf 133}, 229 (2003);
  A. E. Leanhardt, Y. Shin, A. P. Chikkatur, D. Kielpinski,
  W. Ketterle and D. E. Pritchard, Pys. Rev. Lett. {\bf 90}, 100404
  (2003); Y. J. Lin, I. Teper, C. Chin and V. Vuleti{\'{c}},
  Phys. Rev. Lett. {\bf 92}, 050404 (2004).

\bibitem{chan-2001} H. B. Chan, V. A. Aksyuk, R. N. Kleiman,
  D. J. Bishop and F. Capasso, Phys. Rev. Lett. {\bf 87}, 211801
  (2001); H. B. Chan, V. A. Aksyuk, R. N. Kleiman, D. J. Bishop, and
  Federico Capasso, Science {\bf 291}, 1941 (2001).

\bibitem{dobson-2006} J. F. Dobson, A. White, A. Rubio,
  Phys. Rev. Lett. \textbf{96}, 073201 (2006).

\bibitem{hamaker-1937} H. C. Hamaker, Physica (Amsterdam) {\bf 4},
  1058 (1937).

\bibitem{jeziorsky94}  B. Jeziorski, R. Moszynski and K. Szalewicz,
  Chem. Rev. {\bf 94}, 1887 (1994).

\bibitem{lennard-1932} J. E. Lennard-Jones, Trans. Faraday Soc. {\bf
    28}, 333 (1932).

\bibitem{casimir-1946} H. B. G. Casimir and D. Polder, Nature {\bf
    158}, 787 (1946); Phys. Rev. {\bf 73}, 360 (1948).

\bibitem{Lifshitz56} E. M. Lifshitz, Zh. Eksp. Teor. Fiz. \textbf{29},
  94 (1956) [Sov. Phys. JETP \textbf{2} 73, (1956)].

\bibitem{zaremba76} E. Zaremba and W. Kohn, Phys. Rev. B \textbf{13},
  2270 (1976).

\bibitem{sandoghdar-1992} V. Sandoghdar, C. I. Sukenik, E. A. Hinds
  and S. Haroche, Phys. Rev. Lett. {\bf 68}, 3432 (1992).

\bibitem{fichet-2007} M. Fichet {\em et al}, Eur. Phys. Lett. {\bf
    77}, 54001 (2007), and references therein.

\bibitem{caride05} A. O. Caride, G. L. Klimchitskaya,
  V. M. Mostepanenko, and S. I. Zanette, Phys. Rev. A \textbf{71},
  042901 (2005).

\bibitem{patil-2002} S. H. Patil, K. T. Tang and J. P. Toennies,
  J. Chem. Phys. {\bf 116}, 8118 (2002); 
  E. Hult and A. Kiejna, Surf. Sci. {\bf 383},  (1997);
A. Liebsch, Phys. Rev. B {\bf 35}, 9030 (1987);
G. Vidali and M. W. Cole, Surf. Sci. {\bf 110}, 10 (1981).

\bibitem{andersson-1996}  Y. Andersson,
D. C. Langreth, and B. I. Lundqvist, Phys. Rev. Lett. {\bf 76}, 102 (1996);
E. Hult, Y. Andersson, B. I. Lundqvist and D. C. Langreth, Phys. Rev. Lett. {\bf
77}, 2029 (1996); E. Hult, H. Rydberg,
  B. I. Lundqvist and D. C. Langreth, Phys. Rev. B {\bf 59}, 4708 (1999);
Y. Andersson and H. Rydberg, Phys. Scripta {\bf 60}, 211 (1999).

\bibitem{tddft-book}  M. A. L. Marques, C. Ullrich, F. Nogueira, A.
Rubio and E.K.U. Gross (editors), \textit{Time-Dependent
Density-Functional Theory}, Lecture Notes in Physics 706, Springer
Verlag, Berlin (2006).


\bibitem{vanGisbergen-1995} S. J. A. van Gisbergen, J. G. Snijders,
  and E. J. Baerends, J. Chem. Phys. {\bf 103}, 9347 (1995).

\bibitem{marques-vdw-07} M. A. L. Marques, A. Castro, G. Malloci, G.
  Mulas, and S. Botti, J. Chem. Phys. 127, 014107 (2007).

\bibitem{norman01} P. Norman, D. M. Bishop, H. J. A. Jensen and J.
  Oddershede, J. Chem. Phys. {\bf 115}, 10323 (2001); Norman, A.
  Jiemchooroj and B. E. Semelius, J. Chem. Phys. {\bf 118}, 9167
  (2003); A. Jiemchooroj, P. Norman and B. E. Semelius, J. Chem. Phys.
  {\bf 123}, 124312 (2005); A. Jiemchooroj, P. Norman and B. E.
  Semelius, J. Chem. Phys. {\bf 125}, 124306 (2006).

\bibitem{banerjee00} A. Banerjee and M. K. Harbola, J. Chem. Phys.
  {\bf 117}, 7845 (2002); A. Banerjee and M. K. Harbola, PRAMANA J.
  Phys. {\bf 66}, 423 (2006); A. Banerjee and M. K. Harbola,
  arXiv:0801.1424v1 (2008).

\bibitem{sternheimer-1951} R. Sternheimer, Phys. Rev. {\bf 84}, 244
  (1951).

\bibitem{baroni-2001} S. Baroni, S. de Gironcoli, A. D. Corso, and P.
  Gianozzi, Rev. Mod. Phys. {\bf 73}, 515 (2001), and references
  therein.

\bibitem{senatore-1987} G. Senatore and K. R. Subbaswamy, Phys. Rev. A
  {\bf 35}, 2440 (1987); S. P. Karna and M. Dupuis, Chem. Phys. Lett.
  {\bf 171}, 201 (1990); S. J. A. van Gisbergen, J. G. Snijders, and
  E. J. Baerends, Phys. Rev. Lett.  {\bf 78}, 3097 (1997); J. I.
  Iwata, K. Yabana, and G. F. Bertsch, J. Chem. Phys. {\bf 115}, 8773
  (2001); B. Walker, A. M. Saitta, R. Gebauer, and S. Baroni, Phys.
  Rev. Lett. {\bf 96}, 113001 (2006).

\bibitem{octopus}  M. A. L. Marques, A. Castro, G. F. Bertsch and A. Rubio,
  Comp. Phys. Comm. \textbf{151}, 60 (2003);
  A. Castro, M. A. L. Marques, H. Appel, M. Oliveira, C. Rozzi,
  X. Andrade, F. Lorenzen, E. K. U. Gross and A. Rubio,
  phys. stat. sol. (b) \textbf{243}, 2465 (2006).

\bibitem{andrade07} X. Andrade, S. Botti, M. A. L. Marques, and A.
  Rubio, J. Chem. Phys. \textbf{126}, 184106 (2007).

\bibitem{hedin80} L. Hedin, Phys. Scripta \textbf{21}, 477 (1980).

\bibitem{onida-2002} G. Onida, A. Rubio and L. Reining, Rev. Mod.
  Phys. {\bf 74}, 601 (2002).


\bibitem{delsole93} R. Del Sole and R. Girlanda, Phys. Rev. B
  \textbf{48}, 11789 (1993).

\bibitem{hanke78} W. Hanke, Adv. Phys. \textbf{27}, 287 (1978).

\bibitem{botti-review07} S. Botti, A. Schindlmayr, R. Del Sole, and L.
  Reining, Rep. Prog. Phys. \textbf{70}, 357 (2007), and references
  therein.

\bibitem{reining02} L. Reining, V. Olevano, A. Rubio, G. Onida, Phys.
  Rev. Lett. {\bf 88}, 066404 (2002).

\bibitem{botti04} S. Botti, F. Sottile, N. Vast, V. Olevano, L.
  Reining, H. C. Weissker, A. Rubio, G. Onida, R. Del Sole, R. W.
  Godby, Phys. Rev. B {\bf 69}, 155112 (2004).

\bibitem{botti05} S. Botti, A. Fourreau, F. Nguyen, Y. O. Renault, F.
  Sottile and L. Reining, Phys. Rev. B {\bf 72}, 125203 (2005).

\bibitem{dp} V. Olevano, L. Reining and F. Sottile, http://theory.polytechnique.fr/codes/dp/.

\bibitem{siesta} J.\,M. Soler, E. Artacho, J.\,D. Gale, A.
  Garc\'{\i}a, J.  Junquera, P. Ordej\'on, and D. S\'anchez-Portal, J.
  Phys.: Condens. Matter \textbf{14}, 2745 (2002).

\bibitem{pbe96} J.\,P. Perdew, K. Burke, and M. Ernzerhof, Phys. Rev.
  Lett. \textbf{77}, 3865 (1996).

\bibitem{troullier91} N. Troullier and J. L. Martins, Phys. Rev. B
  \textbf{43}, 1993 (1991).

\bibitem{perdew-zunger81} J. P. Perdew and A. Zunger, Phys. Rev. B
  \textbf{23}, 5048 (1981).

\bibitem{vasiliev-2000} I. Vasiliev, S. Ogut, and J.
  R. Chelikowsky, Phys. Rev. Lett. {\bf 86}, 1813 (2001); M. C.
  Troparevsky, L. Kronik, and James R. Chelikowsky, Phys. Rev. B {\bf
    65}, 033311 (2001). S. Botti and M. A. L. Marques, Phys. Rev. B
  {\bf 75}, 035311 (2007).

\bibitem{marques-2001} M. A. L. Marques, A. Castro and A. Rubio, J.
  Chem. Phys. {\bf 115}, 3006 (2001).

\bibitem{abinit} X. Gonze {\em et al}, Comp. Mat. Sci. {\bf 25}, 478
  (2002); X. Gonze {\em et al}, Zeit. Kristallogr. {\bf 220}, 558
  (2005).

\bibitem{hamman} D. R. Hamann, M. Schl{\"{u}}ter and C. Chiang, Phys.
  Rev. Lett. {\bf 43}, 1494 (1979).

\bibitem{wolkenstein41} M. V. Wolkenstein, C. R. Acad. Sci. URSS {\bf 30}, 791 (1941).

\bibitem{cardona82} {\em Light Scattering in Solids II}, edited by M. Cardona and
G. G\"untherodth (Springer-Verlag, Berlin, 1982).

\bibitem{jiemchooroj04} A. Jiemchooroj, B. E. Sernelius, and P. Norman,
  Phys. Rev. A \textbf{69}, 044701 (2004).

\bibitem{grosso-pastori00} G. Grosso and G. Pastori Parravicini, {\em
    Solid State Physics}, Elsevier Academic Press, London (2000).

\bibitem{emt} J. C. Maxwell-Garnett, Philos. Trans. R. Soc. {\bf 203}, 385
(1904); D. M. Wood and N. W. Ashcroft, Philos. Mag. {\bf 35},
269 (1977).

\end{thebibliography}
\end{document}